# Magnetic structure of $Cu_2MnBO_5$ ludwigite: thermodynamic, magnetic properties and neutron diffraction study

Evgeniya Moshkina[1,2], Clemens Ritter[3], Evgeniy Eremin[1,4], Svetlana Sofronova[1], Andrey Kartashev[1,5], Andrey Dubrovskiy[1], and Leonard Bezmaternykh[1]

[1]Kirensky Institute of Physics, Federal Research Center KSC SB RAS, Krasnoyarsk, 660036 Russia
[2]Siberian State Aerospace University, Krasnoyarsk, 660014 Russia
[3]Institut Max von Laue - Paul Langevin, BP 156, F-38042 Grenoble Cedex 9, France
[4]Siberian Federal University, Krasnoyarsk, 660041 Russia
[5]Krasnoyarsk State Pedagogical University, Krasnoyarsk, 660049 Russia

**Abstract**—We report on the thermodynamic, magnetic properties and the magnetic structure of ludwigite-type $Cu_2MnBO_5$. The specific heat, the low-field magnetization and the paramagnetic susceptibility were studied on a single crystal and combined with powder neutron diffraction data. The temperature dependence of the specific heat and the neutron diffraction pattern reveal a single magnetic phase transition at T=92 K, which corresponds to the magnetic ordering into a ferromagnetic phase. The cation distribution and the values and directions of magnetic moments of ions in different crystallographic sites are established. The magnetic moments of $Cu^{2+}$ and $Mn^{3+}$ ions occupying different magnetic sites in the ferrimagnetic phase are pairwise antiparallel and their directions do not coincide with the directions of the principal crystallographic axes. The small value of the magnetic moment of copper ions occupying site 2a is indicative of partial disordering of the magnetic moments on this site. The magnetization measurements show a strong temperature hysteresis of magnetization, which evidences for field-dependent transitions below the phase transition temperature.

## I. Introduction

$Cu_2MnBO_5$ belongs to the family of quasi-two-dimensional oxyborates with the ludwigite structure. Ludwigites have a complex crystal structure, which involves quasi-low-dimensional elements (zigzag walls and three-leggedladders) formed by metal-oxygen octahedral [1–3]. The ludwigite unit cell contains four formula units and includes divalent and trivalent cations or divalent and tetravalent ones. In this structure, metal cations are distributed over four nonequivalent positions.

The complex crystallographic structure and the presence of four nonequivalent positions occupied by magnetic cations lead to the formation of complex magnetic structures in the ludwigite-type crystals. In view of this, it is complex and often impossible to determine the configuration of magnetic moments using macroscopic magnetization measurements. In addition, the ludwigite structure is characterized by the large number of triangular groups formed by metal cations, which sometimes leads to the occurrence of frustrations and spin-glass-like states [2–12].

To date, the microscopic magnetic structure has been experimentally determined only for the monometallic ludwigites $Co_3BO_5$ and $Fe_3BO_5$ [10-12]. An important feature of the $Co_3BO_5$ and $Fe_3BO_5$ ludwigites is the division of the magnetic structure into two subsystems. In $Fe_3BO_5$, which sees a charge ordering transition just below room temperature, the magnetic subsystems order at different temperatures with mutually orthogonal magnetic moments [11]. The $Co_3BO_5$ ludwigite displays a single magnetic transition with the presence of an ordered arrangement of low spin and high spin states of the $Co^{3+}$ ions ($S_{Co3+}=0$) [12]. These features occur most likely to weaken the frustrations in the system. The magnetic structure of ludwigites containing different magnetic cations have not yet been experimentally investigated; however, from the behavior of their physical properties it was concluded that the magnetic ordering could possibly not involve all subsystems and that in some compounds the magnetization of different sublattices could order at different temperatures and point in different directions [4, 13].

The existence of Mn−Cu ludwigites was reported just recently [6]. Single-crystal samples were synthesized and the primary structural and magnetic characterization was performed for the composition Mn:Cu=1:1 ($Cu_{1.5}Mn_{1.5}BO_5$). Similar to other Cu-containing ludwigites, the synthesized compound has a monoclinically distorted ludwigite structure [7]. Due to the presence of quasi-low-dimensional elements in the structure, many ludwigites in the ordered phase are characterized by a strong magnetic anisotropy between the directions H||$c$ and H⊥$c$, where $c$ is the hard magnetization axis [4, 8, 9]. However, in the $Cu_{1.5}Mn_{1.5}BO_5$ ludwigite, the anisotropy is weak and the difference between the magnetic moment values is only M(H||c):M(H⊥c)=1.5. This represents a fundamental difference from other ludwigite-type compounds. In addition, in contrast to other Mn-containing ludwigites, the $Cu_{1.5}Mn_{1.5}BO_5$ compound has a large magnetic moment, which exceeds e.g. tenfold the magnetic moment of $Ni_{1.5}Mn_{1.5}BO_5$ [6].

Here we report on thorough investigations of the physical properties of the $Cu_2MnBO_5$ ludwigite with a different cation ratio. In contrast to the previously investigated $Cu_{1.5}Mn_{1.5}BO_5$ compound, manganese ions in this ludwigite are mainly in the state with valence 3+, which reduces the probability of admixing divalent manganese to the $Cu^{2+}$ ions. In our previous study [5], we synthesized the $Cu_2MnBO_5$ ludwigite single crystals by the flux technique. It was the first study on this compound, where its structural and magnetic properties were investigated; in particular, the composition was refined, the structure was clarified, the magnetic transition temperature was determined, the strong hysteresis in the field-cooling (FC) and zero field-cooling (ZFC) modes was established, and an anomaly in the magnetization curves near 75 K was found. The group theoretical analysis was performed, the indirect exchange interactions were calculated in the framework of the Anderson−Zavadsky model, and a model of the magnetic structure was proposed.

To shed light on the microscopic nature of the magnetic behavior and clarify the mechanisms of the magnetic phase transition, we studied the magnetic structure of the $Cu_2MnBO_5$ ludwigite using powder neutron diffraction, measured and interpreted the temperature dependence of specific heat of the crystal, established orientational field-temperature dependences of magnetization, and analyzed temperature dependences of the magnetic susceptibility.

## II. Experimental Details

The $Cu_2MnBO_5$ ludwigite single crystals were grown by the flux technique. The crystallization conditions were described in detail in [5].

Magnetic measurements of the $Cu_2MnBO_5$ single crystal were performed on a Physical Property Measurements System PPMS-9 (Quantum Design) at temperatures of $T$=3−300 K in magnetic fields of up to 80 kOe.

Specific heat was measured using an original adiabatic calorimeter with three screens at temperatures from ~64 K (slightly below the nitrogen melting point) to ~320 K [14]. At low temperatures (down to 2 K), the measurements were performed ona PPMS facility (Quantum Design). The specific heat determination error was no more than 1% in both cases.

The investigated sample was a crystal set with a total mass of 244.7 mg. Specific heat of the auxiliary elements (heating pad, lubricant, etc.) was determined separately.

Powder neutron diffraction data were recorded at the Institut Laue Langevin, Grenoble, France, on a D2B high resolution powder diffractometer with a neutron wavelength of $\lambda = 1.594$ Å at room temperature. Due to the fact that the sample had been prepared through crushing of single crystals, strong texture effects became visible in the high resolution neutron powder data. This texture had disappeared only after powdering the sample down to a grain size below 100 µm. The sample was placed in a cylindrical double-wall vanadium container in order to reduce the absorption resulting from the B10 isotope. The temperature dependence of the neutron diffraction pattern was measured on a D20 high-intensity powder diffractometer, as well situated at the Institut Laue Langevin, with $\lambda = 2.41$ Å between 1.6 K and 150 K taking spectra of 5 min every degree. Additional data were taken at base temperature (1.6 K) and at 110 K with the longer acquisition time of 45 min. As the absorption of the sample is stronger at $\lambda = 2.41$ Å than at $\lambda = 1.594$ Å, the sample had to be additionally diluted for these measurements by adding aluminum powder. All neutron data were analyzed using the Rietveld refinement program FULLPROF [15]. The aluminum powder was refined as a second phase. Magnetic symmetry analysis was performed using the program BASIREPS [16, 17].

### III. Magnetic Properties

Figure 1 shows the temperature dependences of magnetization of the investigated $Cu_2MnBO_5$ single crystal, which were obtained in the FC (cooling in nonzero magnetic field) and FH (sample heating after precooling in nonzero magnetic field) regimes at H=200 Oe (H||a). At a temperature of $T \approx 90–92$ K, both curves reveal the sharp magnetization growth corresponding to the phase transition from the paramagnetic to the magnetically ordered state. In the vicinity of the phase transition temperature, one can observe a small hysteresis of the FC and FH dependences with a value of $\Delta T_1 \approx 0.8$ K. At lower temperatures, the dependences exhibit an anomalously strong temperature hysteresis in the range of $T \approx 46–85$ K with a value of $\Delta T_2 \approx 14$ K at H=200 Oe. To study this phenomenon, temperature dependences of the magnetization were measured as well in fields of H=20, 50, and 1000 Oe. The measurements show that the width of the hysteresis depends nonlinearly on the applied magnetic field; specifically, at H=50 Oe, we have $\Delta T_2 \approx 5$ K and at H=20 and 1000 Oe, the temperature hysteresis is less than $\Delta T_2 \approx 1$ K.

When measuring the orientational dependences of the sample magnetization, we used a crystal with the natural habit in the form of a quadrangular prism. Magnetization was measured along the *x, y* and *z* geometrical axes of the prism. The *z* axis coincided with the *a* crystallographic axis and the *x* and *y* axes corresponded to the (1 1 0) and (-1 1 0) crystallographic directions.

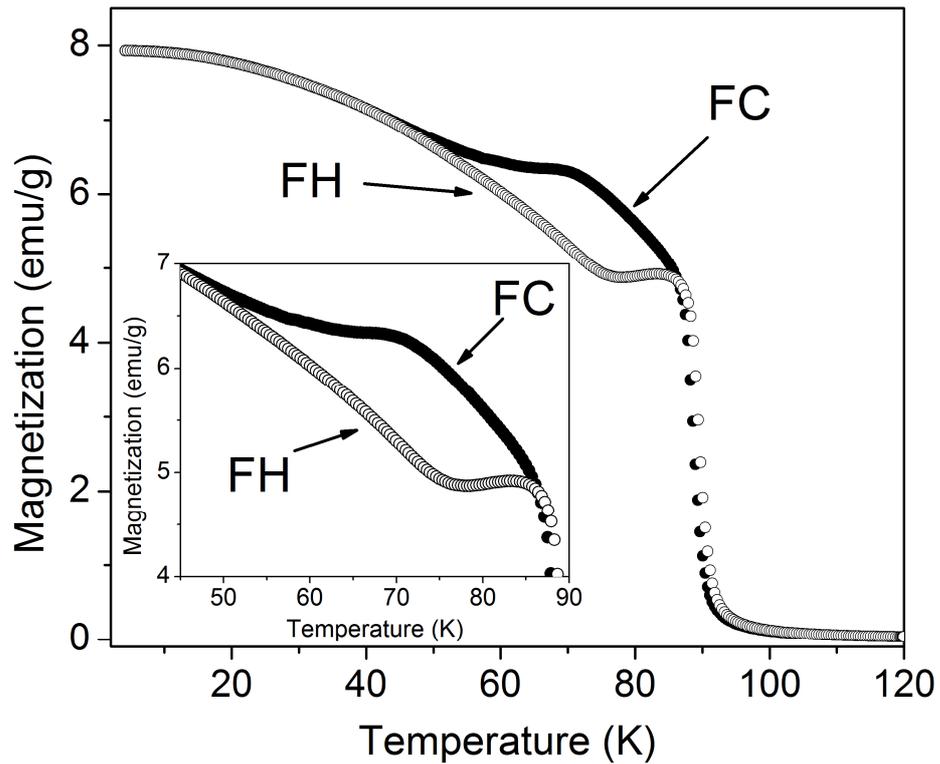

FIG. 1. Temperature dependences of magnetization obtained in the FC (cooling at H=200 Oe) and FH (sample heating in a field of H=200 Oe after precooling at H=200 Oe) regimes (H||c).

Figure 2 presents the orientational dependences of magnetization of the $Cu_2MnBO_5$ sample obtained in a magnetic field of H=1 kOe. All the curves contain the broad asymmetrical maximum, which evidences for the existence of the domain structure in the crystal. The position of this maximum changes depending on the magnetic field direction; in the direction H||x, one can observe a shelf (constant magnetic moment region) in the temperature range of 5−15 K.

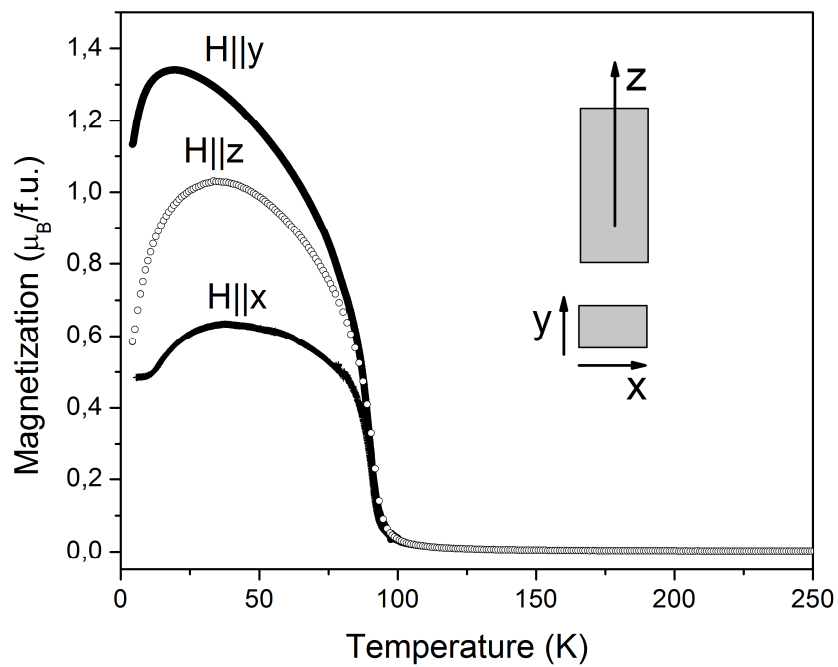

FIG. 2. Temperature dependences of magnetization obtained in a magnetic field of H=1000 Oe applied in the macroscopic directions H||x, H||y, and H||z of the single-crystal samples with the natural habit.

The temperature dependences of the inverse molar susceptibility for H||x, H||y, and H||z in the temperature range of $T=2-300$ K are presented in Figure 3. It can be seen that above the magnetic transition temperature the experimental data obtained with different magnetic field directions do not coincide; i.e., the paramagnetic phase is characterized by anisotropy. This anisotropy can result from the strong g-factor anisotropy caused by the coexistence of two Jahn–Tellerions, $Cu^{2+}$ and $Mn^{3+}$. Not being part of this study of the low temperature behavior we will check this later by studying the electron spin resonance (ESR) spectra.

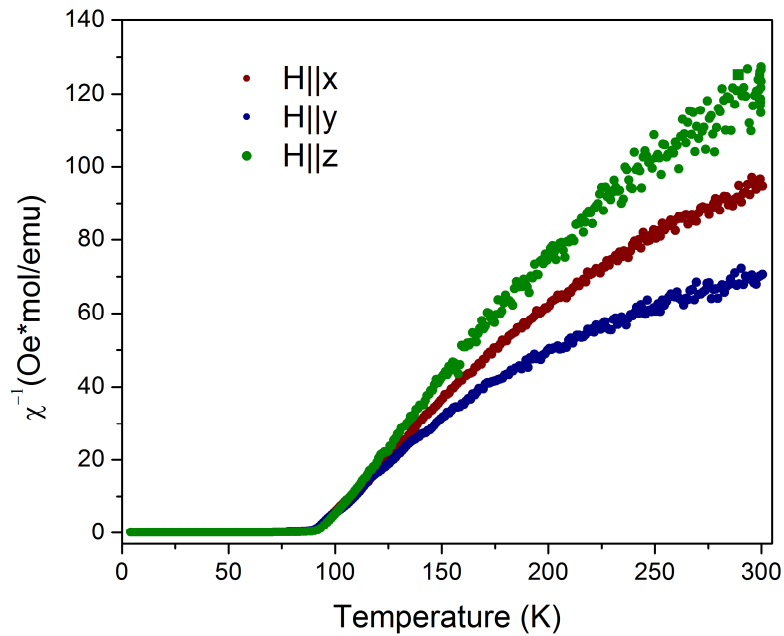

FIG. 3. Inverse susceptibility of $Cu_2MnBO_5$ (H||x, H||y, and H||z).

### IV. Powder Neutron Diffraction

Figure 4 shows the refinement of the high resolution data taken at room temperature. The compound crystallizes in the space group $P2_1/c$ as already proposed by Bezmaternykh et al.[6] for a compound with composition $Cu_{1.5}Mn_{1.5}BO_5$. In this structure, the Mn and Cu cations are distributed over four different sites. Due to the strongly differing neutron scattering lengths for Mn ($b_{coh} = -3.73$ fm) and Cu ($b_{coh} = 7.72$ fm), it is possible to determine precisely the cation distribution over these four sites.

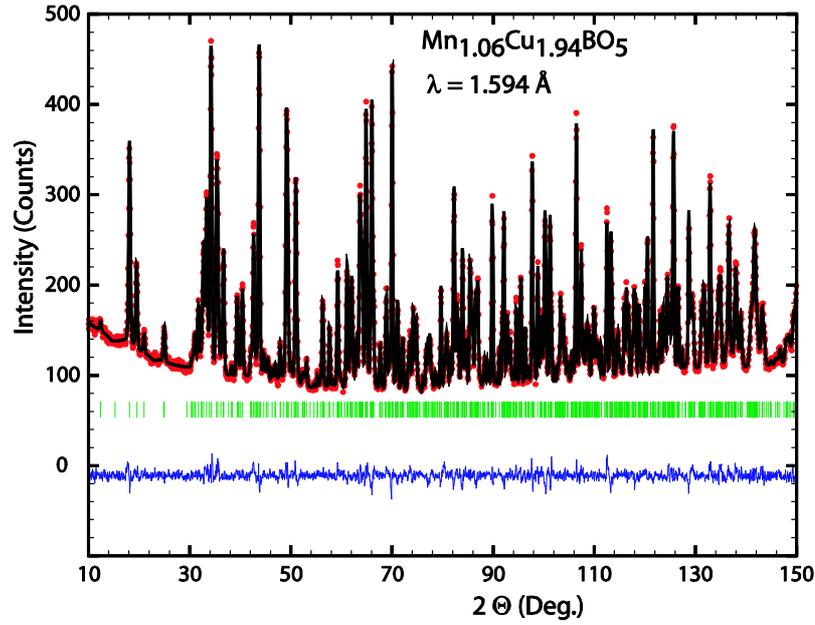

FIG. 4. Observed (dots, red), calculated (black, line), and difference pattern of $Cu_{1.94}Mn_{1.06}BO_5$ at 295 K. The tick marks indicate the calculated position of the nuclear Bragg peaks.

Table 1 gives the lattice parameters, atom coordinates, and the occupations resulting from the refinement. It can be seen that there is a clear site preference with the $Mn^{3+}$ cation occupying almost exclusively one of the $4e$ sites (labelled $4e^2$ in Table 1), while the $Cu^{2+}$ ion is found at a 90% level on the $4e^1$ and the $2d$ and $2a$ sites. The refined stoichiometry corresponds to a $Cu_{1.94(1)}Mn_{1.06(1)}BO_5$ compound. Bond valence calculations using the determined interatomic distances confirm the assumed valences of +3 for Mn and +2 for Cu. This structure is monoclinically distorted with respect to the structure of the closely related $Fe_3BO_5$ compound, which crystallizes in space group $P$bam at room temperature [11]. $Fe_3BO_5$ sees depending on their valence a strong site preference for $Fe^{3+}$ and $Fe^{2+}$ cations: while $Fe^{2+}$ resides on sites $4g$ and $2a$ (space group $P$bam), $Fe^{3+}$ is preferentially found on sites $4h$ and $2d$. This can be compared to the situation in our $Mn_{1.06}Cu_{1.94}BO_5$ compound, where $Mn^{3+}$ is mostly found on site $4e^2$, which corresponds to site $4h$ in $P$bam.

TABLE I. Results of the Rietveld refinement of the high-resolution neutron diffraction data at 295 K for $Cu_{1.94}Mn_{1.06}BO_5$ in $P2_1/c$.

| $P2_1/c$ | X | Y | Z | Occ.$_{Mn/Cu}$ |
|---|---|---|---|---|
| $2a$ | 0 | ½ | ½ | 0.090(4)/0.910(4) |
| $2d$ | ½ | 0 | ½ | 0.068(4)/0.932(4) |
| $4e^1$ | 0.0638(6) | 0.9877(2) | 0.2790(1) | 0.102(4)/0.898(4) |
| $4e^2$ | 0.576(2) | 0.7324(5) | 0.3785(4) | 0.877(2)/0.123(2) |
| B | 0.4057(8) | 0.2640(2) | 0.3670(2) | |
| $O_1$ | 0.0038(8) | 0.0953(2) | 0.1454(2) | |
| $O_2$ | 0.1492(8) | 0.8725(2) | 0.4118(2) | |
| $O_3$ | 0.4661(8) | 0.1187(2) | 0.3654(2) | |
| $O_4$ | 0.6091(8) | 0.6597(2) | 0.5369(2) | |
| $O_5$ | 0.6419(7) | 0.8332(2) | 0.2337(2) | |
| $a$ [Å] | 3.13851(4) | | | |
| $b$ [Å] | 9.4002(1) | | | |
| $c$ [Å] | 12.0204(1) | | | |
| $\beta$ [°] | 92.267(1) | | | |

| $R_{\text{Bragg}}$ | 3.7 | | | |

Figure 5 shows the low-angle region of the thermal dependence of the neutron diffraction pattern of $Cu_{1.94}Mn_{1.06}BO_5$. A transition is clearly visible at about 90 K, where an increase in the intensity of several Bragg reflections can be discerned. In accordance with the magnetic data, this transition is identified as a transition to a magnetically ordered, most probably ferromagnetic state. Down to the lowest temperatures, there is no further transition.

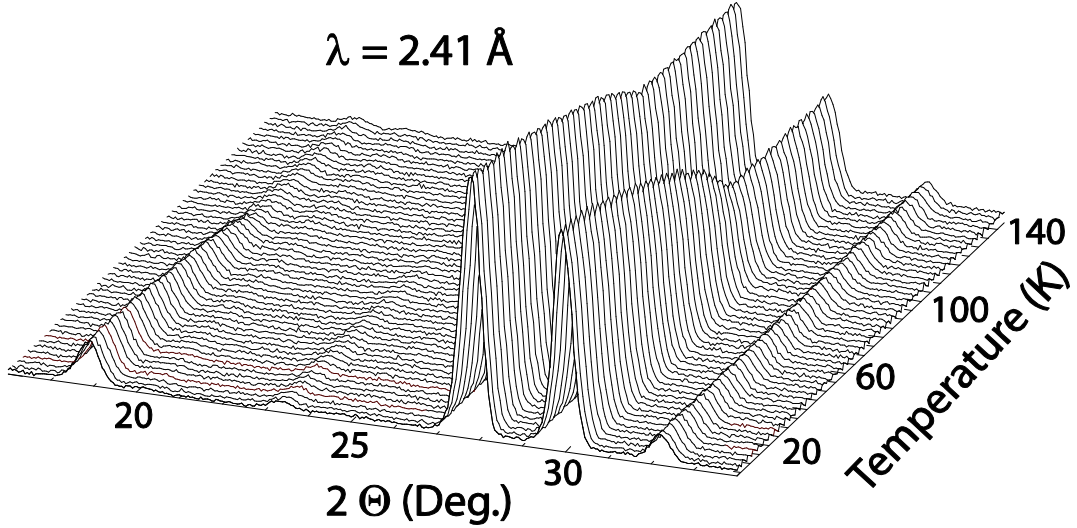

FIG. 5. Thermal dependence of the neutron diffraction pattern of $Cu_{1.94}Mn_{1.06}BO_5$ between 2 K and 140 K. Only every third spectrum of the original measurement is shown.

Using the program K-search, which is a part of the FULLPROF suite of refinement programs, the magnetic propagation vector $\kappa = 0$ was confirmed. Fitting the intensity of the Bragg peak having the most intense magnetic contribution, a transition temperature of $T_C = 92$ K was established. Magnetic symmetry analysis using the program BASIREPS was used to determine for $\kappa = 0$ the allowed irreducible representations (IR) and their basis vectors (BV) for cation sites 4e, 2d, and 2a; they are listed in Table 2.

TABLE II. Basis vectors (BV) of the allowed irreducible representations (IR) for $\kappa = 0$ for the Wykoff positions 4e, 2d and 2a of space group $P2_1/c$

| | IR1 | | | IR2 | | | IR3 | | | IR4 | | |
|---|---|---|---|---|---|---|---|---|---|---|---|---|
| 4e | BV1 | BV2 | BV3 | BV1 | BV2 | BV3 | BV1 | BV2 | BV3 | BV1 | BV2 | BV3 |
| x, y, z | 1 0 0 | 0 1 0 | 0 0 1 | 1 0 0 | 0 1 0 | 0 0 1 | 1 0 0 | 0 1 0 | 0 0 1 | 1 0 0 | 0 1 0 | 0 0 1 |
| -x, y+½, -z+½ | -1 0 0 | 0 1 0 | 0 0 -1 | -1 0 0 | 0 1 0 | 0 0 -1 | 1 0 0 | 0 -1 0 | 0 0 1 | 1 0 0 | 0 -1 0 | 0 0 1 |
| -x, -y, -z | 1 0 0 | 0 1 0 | 0 0 1 | -1 0 0 | 0 -1 0 | 0 0 -1 | 1 0 0 | 0 1 0 | 0 0 1 | -1 0 0 | 0 -1 0 | 0 0 -1 |
| x, -y+½, z+½ | -1 0 0 | 0 1 0 | 0 0 -1 | 1 0 0 | 0 -1 0 | 0 0 1 | 1 0 0 | 0 -1 0 | 0 0 1 | -1 0 0 | 0 1 0 | 0 0 -1 |
| | | | | | | | | | | | | |
| 2d, 2a | | | | | | | | | | | | |
| x, y, z | 1 0 0 | 0 1 0 | 0 0 1 | | | | 1 0 0 | 0 1 0 | 0 0 1 | | | |
| x, -y+½, z+½ | -1 0 0 | 0 1 0 | 0 0 -1 | | | | 1 0 0 | 0 -1 0 | 0 0 1 | | | |

For the determination and refinement of the magnetic structure, a difference data set created by subtracting the high intensity data set taken with long counting times within the paramagnetic phase at 110 K from the data set at 1.6 K was used. This allows refining solely the magnetic contribution and increases thereby the precision of the magnetic moment determination. The fixed scalefactor needed for performing this type of purely magnetic refinement gets first evaluated from the refinement of the 110 K data set. Atomic positions were fixed to the values resulting from the refinement of the high-resolution refinement (Table 1). Testing all the allowed IRs, it is found that the magnetic structure sees a ferromagnetic alignment of spins along the *a* and *c* unit cell directions corresponding to IR3, which corresponds to the one proposed already in [5].There is no contribution coming from BV2 of this IR3, there is therefore no antiferromagnetic component present in the magnetic structure. Figure 6 shows the results of the refinement or the difference data set 2 K – 110 K.

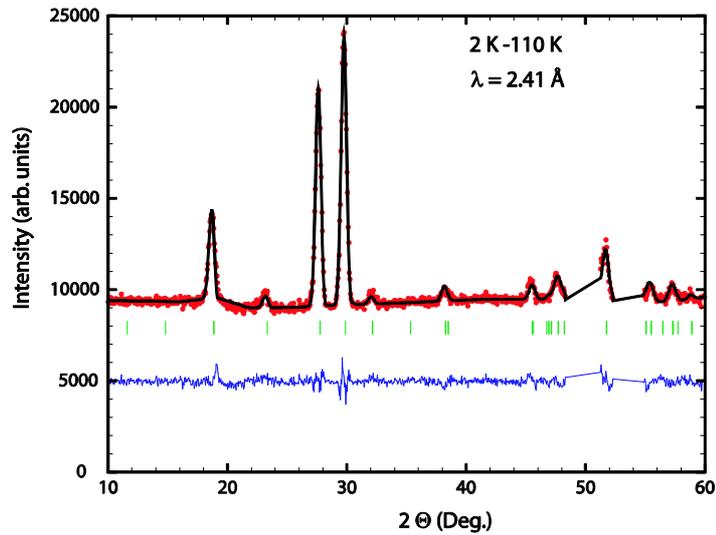

FIG. 6. Refinement of the difference spectrum 2 K – 110 K of $Cu_{1.94}Mn_{1.06}BO_5$. Observed (dots, red), calculated (line, black), and difference pattern. The tick marks indicate the calculated positions of the magnetic Bragg peaks. Two regions at 2 θ ~ 50° and ~ 54° were excluded due to the presence of strong up/down features at the positions nuclear Bragg peaks of the added Al – phase.

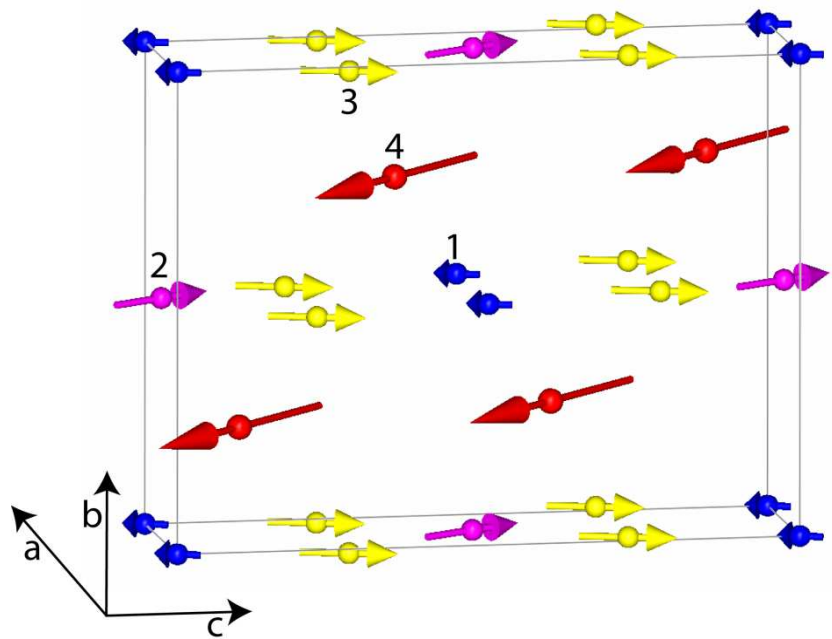

FIG. 7. Magnetic structure of $Cu_{1.94}Mn_{1.06}BO_5$ at 2 K; the numbers correspond to the different cation sites: $4e^2$ mainly occupied by Mn (4), $4e^1$ mainly occupied by Cu (3), $2d$ (2) and $2a$ (1) as well both mainly occupied by Cu.

While the $4e^2$ site, which is mainly occupied by manganese, possesses a magnetic moment of about 2.7 μ$_B$, the $4e^1$ site and the $2d$ and $2a$ sites, which are mainly occupied by copper, have – as expected for a $Cu^{2+}$ ion – lower moment sizes of about 0.9, 1.1 and 0.4 μ$_B$, respectively. The spin directions on the different sites are not parallel, but form an arrangement comprising strong ferrimagnetic elements. Figure 7 displays the magnetic structure where the lengths of the arrows reflect the relative size of the magnetic moments. Table 3 gives details of the refined magnetic components. The corresponding Shubnikov or magnetic space group was determined to P2$_1$'/c' using the programs of the Bilbao Crystallographic Server and of the Isotropy software package [18, 19].

TABLE III. Results of the refinement of the magnetic structure using BV1 and BV3 of IR3. Magnetic components were determined using the $Mn^{3+}$ and the $Cu^{2+}$ magnetic formfactors for the different cation sites depending on which cation occupies predominantly the concerned site. The total magnetic moments μ$_{Tot}$ are given in μ$_B$. The numbering corresponds to the one used in Figure 7 and in the main text.

|  | BV1 | BV3 | μ$_{Tot.}$ |
|---|---|---|---|
| (1) Cu on $2a$ | 0.09(8) | -0.44(9) | 0.45(10) |
| (2) Cu on $2d$ | 0.60(8) | 0.97(6) | 1.12(9) |
| (3) Cu on $4e^1$ | -0.23(3) | 0.91(5) | 0.93(6) |
| (4) Mn on $4e^2$ | -1.93(2) | -1.91(6) | 2.66(6) |
| R$_{Magn.}$ | 5.3 |  |  |

The four different sublattices only possess ferromagnetic interactions, a fact which can be directly linked to the site specific occupation by either $Mn^{3+}$ or $Cu^{2+}$ ions. 90° superexchange interactions M-O-M should in fact be ferromagnetic between cations of the same type having the same valence following the Goodenough–Kanamouri [20] rules. The reduced value of the magnetic moment found for $Mn^{3+}$ - 2.7 μ$_B$ instead of the theoretical 4.0 μ$_B$ – can be related to the non-negligible amount of $Cu^{2+}$ (12%) occupying the $4e^2$ site which will hinder an equivalent amount of neighboring $Mn^{3+}$ cations to adopt a ferromagnetic alignment and could even lead locally to some antiferromagnetic $Mn^{3+}$ - $Cu^{2+}$ interactions.

**V. Thermodynamic Properties**

Figure 8 illustrates the specific heat measurements in the entire temperature rangeinzero magnetic field (T=2–320 K, H=0). One can observe an anomalous behavior with a temperature peak at T$_c$=88.1 K. The lattice specific heat was determined using linear combinations of the Debye–Einstein functions with the characteristic temperatures found to be T$_D$= 331 K and T$_E$= 780 K. It can be seen that the low temperature region isnot correctly interpolated. The same behavior was previously observed in another ludwigite crystal, $Ni_5GeB_2O_{10}$ [13]. Subtracting the lattice contribution to the specific heat from the experimental data, we found the excess specific heat and the phase transition entropy ΔS= 0.6 J/(mol*K). Under the assumption that the magnetic moments order completely in the crystal, the maximum possible entropy of the magnetic phase transition can be calculated from the formula:

$$\Delta S = \Delta S_{Mn} + \Delta S_{Cu} = n_{Mn} R \ln(2S(Mn^{3+})+1) + n_{Cu} R \ln(2S(Cu^{2+})+1) = 25.2 \, J/(mol \cdot K) \quad (1)$$

Where $n_{Mn}$ and $n_{Cu}$ are the ion concentrations, $S(Mn^{3+})=2$ and $S(Cu^{2+})=1/2$ are the spin magnetic moments of ions, and $R$ is the universal gas constant. The magnetic phase transition entropy obtained using formula (1) exceeds by far the experimental value. This difference is indicative of the absence of complete ordering of the magnetic moments at this magnetic phase transition, which agrees with the resultsfrom the neutron magnetic scattering data.The partial ordering of the magnetic moments is characteristic of heterometallic ludwigites, which contain two or more magnetic ions [2, 3]. The homometallic ludwigites $Fe_3BO_5$ [10, 11, 21] and $Co_3BO_5$ [12, 21] are characterized, on the contrary, by the long-range magnetic order.

In addition, we studied the temperature dependence of specific heat in an external magnetic field of H=4.7 kOe (inset a of Figure 8). It can be noted that the temperature of the magnetic phase transition changes only weakly in the applied magnetic field while the specific heat peak is significantly spread. A similar behavior was observed on the completely magnetically ordered ludwigites $Co_3BO_5$[21] and $Co_5SnB_2O_{10}$ [22]. This behavior is indicative of the presence of antiferromagnetic interactions in the crystal [21].

We attribute the anomaly of the excess specific heat at T=23 K (inset b of Figure 8) to additional contributions to the lattice specific heat, which are ignored in the Debye–Einstein models. Although the compound under study is dielectric, at temperatures close to zero, the specific heat decreases in accordance with the linear law, which were observed for all investigated ludwigites [13, 22].

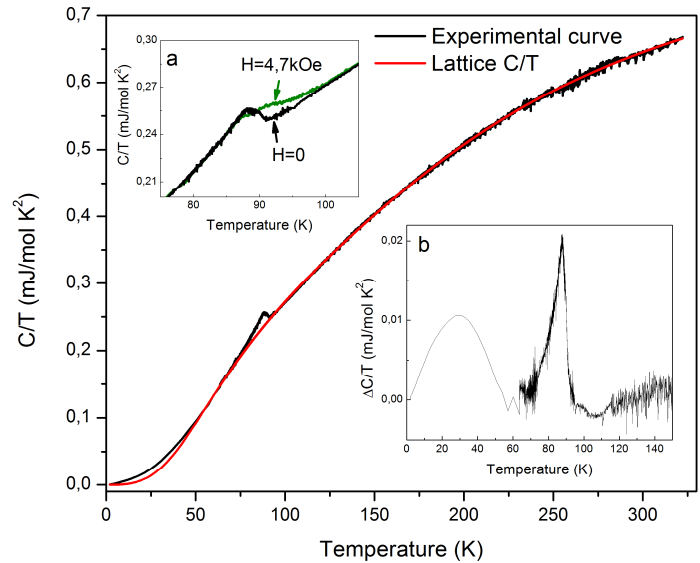

FIG. 8. Specific heat curves (H=0). The black line shows experimental data and the red line, the lattice contribution to specific heat. Inset (a): specific heat curves at H=0 and H=4.7 kOe. Inset (b): residual specific heat.

In Section III, devoted to the magnetic properties of the investigated ludwigite, we found a temperature hysteresis of the magnetization in the heating and cooling modes in magnetic fields of up to H=1 kOe. The dependences of magnetization contain inflection points below the phase transition temperature. To study this effect, we calculated the temperature dependences of the temperature derivative of the squared magnetization (Figure 9), since, according to the molecular field theory, the magnetic contribution to the specific heat is proportional to the squared spontaneous magnetization [23].

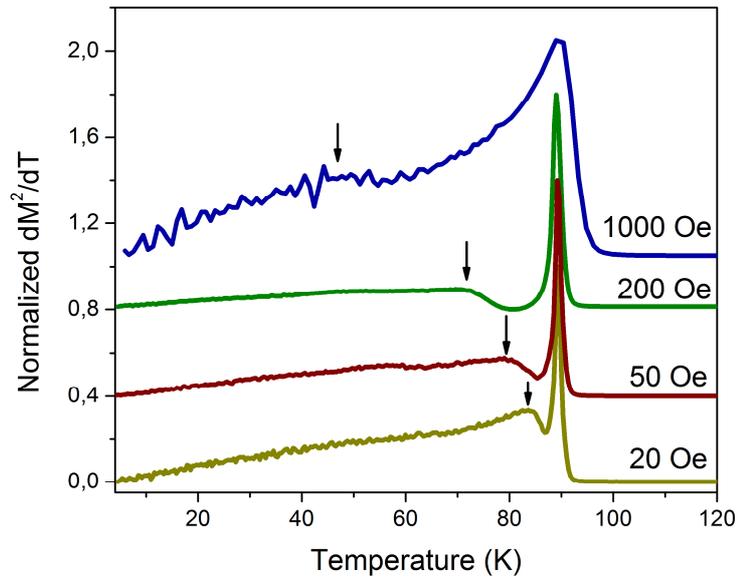

FIG. 9. Temperature dependences of the normalized temperature derivative of the squared magnetization at H=20, 50, 200, and 1000 Oe.

Figure 9 shows the $dM^2/dT(T)$ dependences obtained at H=20, 50, 200, and 1000 Oe. All curves show a peak independent of the external magnetic field which corresponds to the magnetic phase transition at T ≈ 89 K. This temperature is consistent with the phase transition temperature determined from the specific heat measurements and with the neutron diffraction data. However, below the transition temperature T≈89 K, the $dM^2/dT$ (T) dependences show a second peak, whose position and shape depend, to a great extent, on the applied magnetic field. As the magnetic field H is increased, the peak significantly spreads and shifts to lower temperatures.

According to the neutron diffraction data obtained, the $Cu_2MnBO_5$ ludwigite undergoes the only magnetic phase transition at a temperature of $T_c$≈92 K. However, the neutron scattering experiment was carried out at H=0 and, according to the temperature behavior of the derivative of the squared magnetization, in magnetic fields close to zero we can expect the coincidence of the position of the second peak with the phase transition temperature.

The inset in Figure 8 shows the temperature dependence of specific heat in the range of T=82–96 K, which involves the phase transition region. It can be seen that the specific heat peak is fairly broad even without external magnetic field (according to the temperature dependence of the excess specific heat, the peak width attains ΔT≈15 K), which can suggest, e.g., the gradual partial ordering of the moments in the 2a site, which manifests itself as a hysteresis in the magnetization curves.

The dependence of specific heat obtained at H=5 kOe also does not exclude such an interpretation due to the large field value. It can be seen in Figure 9 that at H=1 kOe, the maximum of the derivative significantly broadens and shifts toward lower temperatures. In other words, according to the temperature extrapolation of the center position and peak shape, in a magnetic field of H=5 kOe this peak can be absent.

Such a field dependence of the temperature anomaly peak position is observed in systems with the spin-reorientation transition (see, for example, [24]). As the magnetic field is increased, the temperature of spin reorientation lowers.

## VI. Discussion

To date, the magnetic structure has been determined only for monometallic ludwigites $Co_3BO_5$ [12] and $Fe_3BO_5$ [10, 11]. The results obtained by [10] and [11] for $Fe_3BO_5$ are somewhat different, but the main peculiarities are identical: the magnetic system is divided in two subsystems where the first one comprises the Fe ions on sites 4$h$ and 2$d$ while the second one those of the Fe ions on sites 4$g$ and 2$a$ (*Pbam* setting). The two subsystems form two different three leg ladders (3LL) [4] which order in $Fe_3BO_5$ at different temperatures in perpendicular directions [11]. In the case of $Co_3BO_5$, the magnetic system is as well divided into the same two subsystems which order, however, at the same temperature [11]. In $Fe_3BO_5$ the magnetic moments are directed along the *c* axis in the first subsystem formed by the triad 4-2-4 and along the *b* axis in the second subsystem formed by the triad 3-1-3 [25]. In the $Co_3BO_5$ ludwigite the magnetic order of the 3-1-3 subsystem is the same as in the $Fe_3BO_5$ ludwigite. However the second subsystem 4-2-4, unlike the $Fe_3BO_5$, has almost the same direction as the 3-1-3 subsystem. But, formed by 4-2-4 triads, this 3LL consists only of the chains of the position 2 ions due to the nonmagnetic low spin state of the $Co^{3+}$ ions positioned on site 4. These chains are connected with the 3-1-3 3LL by super-superexchange interactions Co-O-B-O-Co.

In the compound investigated by us, the magnetic moments lie in a different plane– *ac*. However, there is a certain similarity with the magnetic structure of $Fe_3BO_5$ [11]. Figure 10 shows the magnetic moments of ions on each crystallographic site; for convenience, they have a common reference point. It can be seen that the magnetic moments of ions in positions 2 and 4 and in positions 3 and 1 lie almost in one straight and are antiferromagnetically oriented. The two straights make an angle of 60°. Thus, in the crystal under study, similar to $Fe_3BO_5$, the magnetic subsystem is divided in the same two subsystems, but the angle between the magnetic moments amounts to about 60° with the moments lying in the *ac* plane.

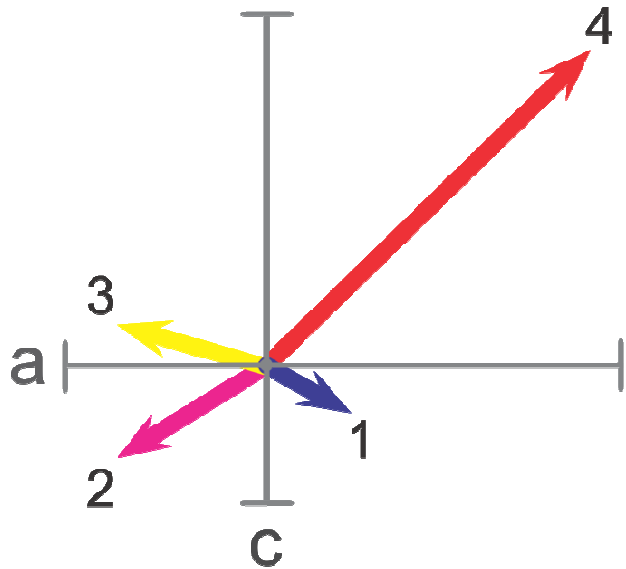

FIG. 10. Orientations of the magnetic moments (NPD data).

The difference in the orientation of the magnetic moments in $Cu_2MnBO_5$ and in $Fe_3BO_5$ can be caused by the Jahn−Teller effect; as mentioned in [5], in $Fe_3BO_5$ the long axes of the oxygen octahedra surrounding iron lie in the *bc* plane, while in $Cu_2MnBO_5$ the octahedra are distorted due to the Jahn–Teller effect and the long axes are turned in the *a* direction as well.

The estimation of the exchange interactions using the Anderson–Zavadsky model shows that in $Fe_3BO_5$ there are many frustrating interactions, since the metal ions in the ludwigite structure form triangular groups and most of them couple in triads with each other [26, 27]. The magnetic moments of the two subsystems arrange orthogonally, possibly, to reduce the frustrations [26]. In $Cu_2MnBO_5$, part of the exchange interactions between the subsystems is also frustrated and the other are very weak [5], which leads, as in $Fe_3BO_5$, to the nonparallel orientation of the moments in the subsystems.

Such a separation of the magnetic system in two subsystems oriented nonparallel is apparently characteristic of all ludwigites; however, up to now the magnetic structure have been only studied for $Fe_3BO_5$, $Co_3BO_5$ and now $Cu_2MnBO_5$. This idea is in directly confirmed by investigations of the magnetization of single crystals of $FeCo_2BO_5$ and $Ni_5GeB_2O_{10}$ [4, 13], which also evidence the occurrence of magnetization in two directions.

One more specific feature of is the small magnetic moment of a copper ion in site 1 ($2a$). The calculation of exchange interactions showed that the exchange interactions with ions in site 4 ($4e^2$) are weakly antiferromagnetic and the exchange interactions with the two nearest ions on site 3 ($4e^1$) are different: one is weakly ferromagnetic and the other, antiferromagnetic.

At the magnetic phase transition, ions in site 1 ($2a$) are apparently weakly coupled by the exchange interaction with the rest ions and order incompletely. The FH and FC temperature dependences of magnetization reveal the above-discussed hysteresis, which can be related to the incomplete ordering of the magnetic moments of ions in site 1 ($2a$) and, as we stated above, the behavior of specific heat does not contradict the proposed model.

### VII. Conclusions

The structural, magnetic, and thermodynamic properties of the ludwigite $Cu_2MnBO_5$, a new compound in the family of quasi-low-dimensional oxyborates with the ludwigite structure, have been studied. The quasi-two-dimensional crystal structure and the presence of a large number of magnetic ions on different sites in the unit cell lead to a magnetic structure which is difficult to establish by macroscopic magnetic studies. The $Cu_2MnBO_5$ ludwigite is the first heterometallic representative of the family of ludwigites whose microscopic magnetic structure was experimentally determined by neutron powder diffraction. Similar studies had been carried out earlier for the monometallic $Fe_3BO_5$ and $Co_3BO_5$ ludwigites. Combining the new results on $Cu_2MnBO_5$ with the results on $Fe_3BO_5$ and $Co_3BO_5$ it appears as a common feature of the ludwigites that the magnetic structure is divided into two subsystems of three leg ladders labelled 4-2-4 and 3-1-3 where the numbers represent the different magnetic cation sites forming the ladders. This characteristic of the magnetic structure is linked to the specific geometry of the crystal structure and occurs to weaken the frustration in the system. The magnetic structure of $Cu_2MnBO_5$ is more complex than in $Fe_3BO_5$ – the directions of all the four magnetic moments do not coincide with the principal crystallographic directions in the crystal, which is most likely caused by the Jahn−Teller effect. In addition, the small moment of the copper ions in site 1 ($2a$) indicates the incomplete magnetic ordering on this site, which is confirmed by the magnetic (anomaly in the magnetization curves) and thermodynamic properties and is characteristic of heterometallic ludwigites [2, 4, 7, 28]. The strong dependence of the magnetization on the applied magnetic field in the region of the second anomaly in the temperature dependences of magnetization needs further investigations of the magnetic and thermodynamic properties in weak magnetic fields. A clear understanding of the mechanisms of magnetic ordering in the $Cu_2MnBO_5$ ludwigite will elucidate the properties of other compounds in this family.

**Acknowledgement**

This study was supported by Russian Foundation for Basic Research (RFBR) and Government of Krasnoyarsk Territory according to the research project No. 16-42- 243028.